\documentclass[aps,prl,twocolumn,showpacs,superscriptaddress]{revtex4}

\usepackage{graphicx}
\usepackage{dcolumn}
\usepackage{bm}
\usepackage{amssymb}
\usepackage{natbib}
\usepackage{amsmath}

\begin{document}

\title{Laser cooling and slowing of CaF molecules}

\author{V. Zhelyazkova}
\author{A. Cournol}
\author{T. E. Wall}
\author{A. Matsushima}
\author{J. J. Hudson}
\author{E. A. Hinds}
\author{M. R. Tarbutt}
\author{B. E. Sauer}

\affiliation{Centre for Cold Matter, Blackett Laboratory, Imperial College London, Prince Consort Road, London SW7 2AZ, United Kingdom}

\begin{abstract}
We demonstrate slowing and longitudinal cooling of a supersonic beam of CaF molecules using counter-propagating laser light resonant with a closed rotational and almost closed vibrational transition. A group of molecules are decelerated by about 20\,m/s by applying light of a fixed frequency for 1.8\,ms. Their velocity spread is reduced, corresponding to a final temperature of about 300\,mK. The velocity is further reduced by chirping the frequency of the light to keep it in resonance as the molecules slow down.
\end{abstract}

\maketitle

There is currently great interest in cooling and controlling molecules, motivated by a wide range of applications \cite{Carr(1)09}. Polar molecules interact through strong, long-range, anisotropic, and controllable dipole-dipole interactions, and so a gas of molecules, at low temperature and under precise control, could be used as a simulator of strongly-interacting quantum systems \cite{Goral02,Micheli06}, or for quantum computation \cite{DeMille02,Andre06}. Molecules are already used in several tests of fundamental physics, such as the measurement of the electron's electric dipole moment \cite{Hudson(1)11, Eckel(1)13, Vutha(1)10}, searches for changes in fundamental constants \cite{Hudson(1)06, Truppe(1)13}, and tests of parity violation \cite{DeMille(1)08, Darquie(1)10}. The precision of these measurements can be improved by cooling the molecules to low temperatures \cite{Tarbutt(1)09, Tar13a}. The availability of cold molecules will also open many new possibilities to study and control chemistry at low temperatures \cite{Balakrishnan01,Krems08}. Ultracold molecules have been produced by binding together ultracold atoms, either by photoassociation \cite{Sage(1)05} or magnetoassociation \cite{Ni(1)08, Danzl(1)10}. Other molecules have been cooled to about 1\,K using a cold buffer gas \cite{Weinstein(1)98}, and beams of molecules have been decelerated and trapped using electric \cite{Bethlem(1)99, Bethlem(1)00, vanVeldhoven(1)05}, magnetic \cite{Vanhaecke(1)07, Narevicius(1)08, Sawyer(1)07} and optical \cite{Fulton(1)06} fields. Once trapped, evaporative cooling to lower temperatures has been demonstrated \cite{Stuhl(1)12}, and sympathetic cooling has been studied \cite{Parazzoli(1)11, Tokunaga(1)11}. Following an initial demonstration of the radiative force acting on SrF \cite{Shuman(1)09}, transverse laser cooling was applied to beams of SrF \cite{Shuman(1)10} and YO \cite{Hummon(1)13} molecules. Radiation pressure has been used to slow down beams of SrF \cite{Barry(1)12}, but longitudinal cooling of a molecular beam has not previously been shown. Here we report longitudinal laser cooling and slowing of a supersonic beam of CaF molecules.

\begin{figure}[b]
\centering
\includegraphics[scale =0.48]{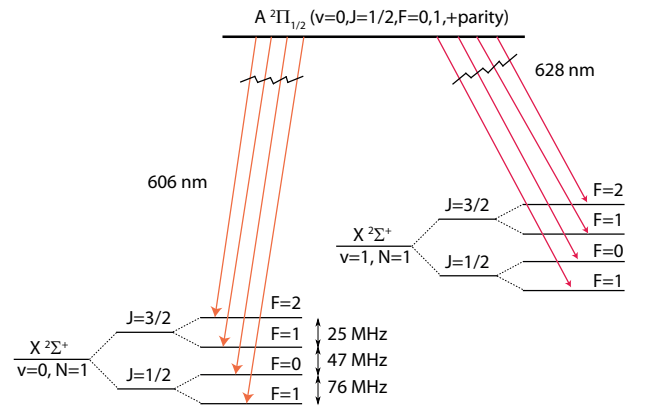}
\caption{(Color online) Laser cooling transitions. The main cooling cycle takes place on the $\text{A}(v'=0)\leftrightarrow \text{X}(v''=0)$ transition, while a vibrational repump laser acts on the $\text{A}(v'=0)\leftrightarrow \text{X}(v''=1) $ transition.}
\label{hyperfineScheme}
\end{figure}

Figure \ref{hyperfineScheme} shows the cooling scheme. We use $v$ to label the vibrational quantum number, and $N$, $J$ and $F$ for the rotational, total electronic, and total angular momentum quantum numbers. We excite the lowest-lying vibrational and rotational state of positive parity in the first electronically excited state, $\text{A} ^{2}\Pi_{1/2} (v'=0,J'=1/2)$, whose lifetime is 19.2\,ns \cite{WallFC}. Here, the hyperfine interaction results in two levels with $F=0$ and 1, separated by approximately $4.8$\,MHz \cite{WallFC}, too small to be resolved in the experiment. Due to the angular momentum and parity selection rules for electric dipole transitions, the population can only decay to the states $\text{X} ^{2}\Sigma^{+} (v'', N''=1)$. There is no restriction on the allowed values of $v''$, but the relative probabilities are governed by the Franck-Condon factors which are approximately 97\% for $v''=0$, 3\% for $v''=1$, 0.08\% for $v''=2$, and negligibly small for all $v''>2$ (see \cite{WallFC, pelegrini} and the present paper). The spin-rotation and hyperfine interactions split each of these states into four components, separated by radio-frequency intervals, as shown in Fig.\,\ref{hyperfineScheme}. We use two lasers, each modulated to provide the four frequencies needed to drive the transitions from $v''=0$ and $v''=1$, labelled $v_{00}$ and $v_{10}$. From this laser light, a molecule scatters 1000-2000 photons on average, before it decays to $v''=2$. Since each photon absorption reduces the velocity by $h/(M \lambda)$=0.011 m/s ($M$ is the molecular mass, $\lambda$ the laser wavelength), substantial changes in velocity are possible.

\begin{figure}[t]
\centering
\includegraphics[scale=0.35]{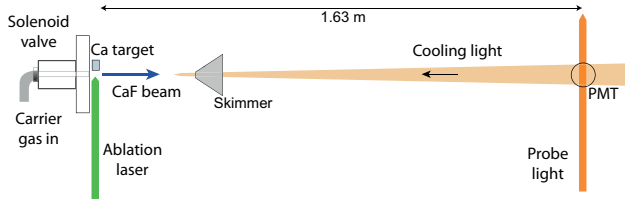}
\caption{(Color online) Schematic of the experimental setup.}
\label{fig:schematic}
 \end{figure}

Figure \ref{fig:schematic} shows the experimental setup. A pulsed beam of CaF molecules is produced by laser ablation of a Ca target into a supersonically expanding pulsed jet of Ar and SF$_6$ \cite{WallFC, Tarbutt(1)02}. The target is at $z=0$, and the ablation laser fires at $t=0$. The pulses have a mean velocity of 600\,m/s and a translational temperature of 3\,K. The beam passes through a 2\,mm diameter skimmer at $z=70$\,mm. At $z=L=1.63$\,m, molecules in any of the $\text{X} ^{2}\Sigma^{+} (v''=0,1,2; N''=1)$ states can be detected by laser-induced fluorescence using a single-frequency probe laser beam that intersects the molecular beam at right angles. The fluorescence signal is recorded with a time resolution of 10\,$\mu$s, giving the time-of-flight (ToF) profile of each molecular pulse. Between pulses, the probe laser frequency is stepped by approximately 0.6\,MHz, so that over a sequence of pulses the frequency is scanned over the four hyperfine components of the P(1) rotational line of the $\text{A} ^{2}\Pi_{1/2} (v) - \text{X} ^{2}\Sigma^{+} (v)$ transition, with $v=0$, 1 or 2 selected by the choice of laser wavelength. We refer to this transition as $\text{A} - \text{X} (v-v)$

 \begin{figure}[b]
\centering
\includegraphics[width=0.45\textwidth]{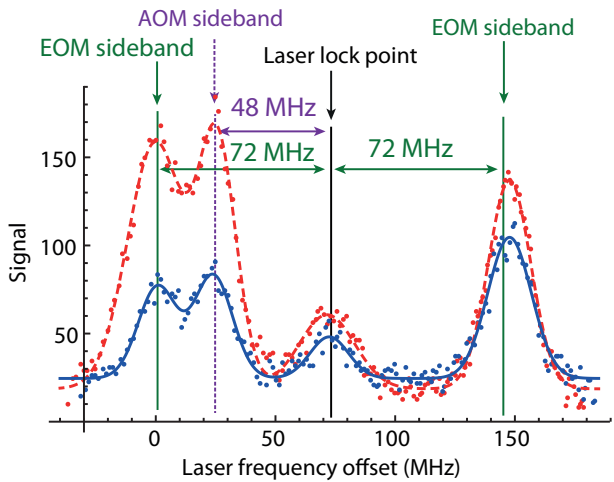}
\caption{(Color online) Spectrum showing the hyperfine structure of the $\text{A} - \text{X} (0-0)$ transition both with (solid blue line) and without (dashed red line) the cooling lasers applied. Dots are experimental data and the lines are fits to four Gaussians. The scheme for generating the frequency sidebands using acousto- and electro-optic modulators is shown.}
\label{spectrum}
\end{figure}

Two cw dye lasers generate the laser cooling light, which counter-propagates to the molecular beam. The rf sidebands addressing the various hyperfine transitions are generated using a combination of an acousto-optic modulator (AOM) and an electro-optic modulator (EOM) as shown in Fig.\,\ref{spectrum}. An additional AOM used as a fast switch turns on the $v_{00}$ cooling light at $t_{\text{start}}=200$\,$\mu$s for a variable duration, $\tau$. The laser beams are then spatially overlapped and double-pass through a final AOM which applies a frequency chirp, $\beta$, to all frequencies simultaneously. This chirp is used to compensate for the changing Doppler shift of the molecules as they slow down. The overlapped beams are shaped to an approximately Gaussian intensity distribution with $1/e^2$ radii of 3.81\,mm at the position of the detector and 0.78\,mm at the skimmer. The total power is 32\,mW, divided approximately equally between the eight frequencies. A mechanical shutter blocks the $v_{00}$ beam on alternate shots of the experiment, so that successive shots record the change in the ToF profile when the cooling is applied, minimizing sensitivity to slow drifts in the molecular flux. The $v_{10}$ light is always applied continuously so that population in $v''=1$ is transferred to $v''=0$. To avoid optically pumping molecules into Zeeman sub-states that do not couple with the linearly polarized light, we apply a $\approx$10\,G magnetic field orthogonal to the molecular beam and at 45$^\circ$ to the laser polarization \cite{Berkeland2002, Shuman(1)09}. We study the slowing and cooling of the molecules as a function of cooling time, $\tau$, and frequency chirp, $\beta$.

\begin{figure}[t]
\centering
\includegraphics[width=0.5\textwidth]{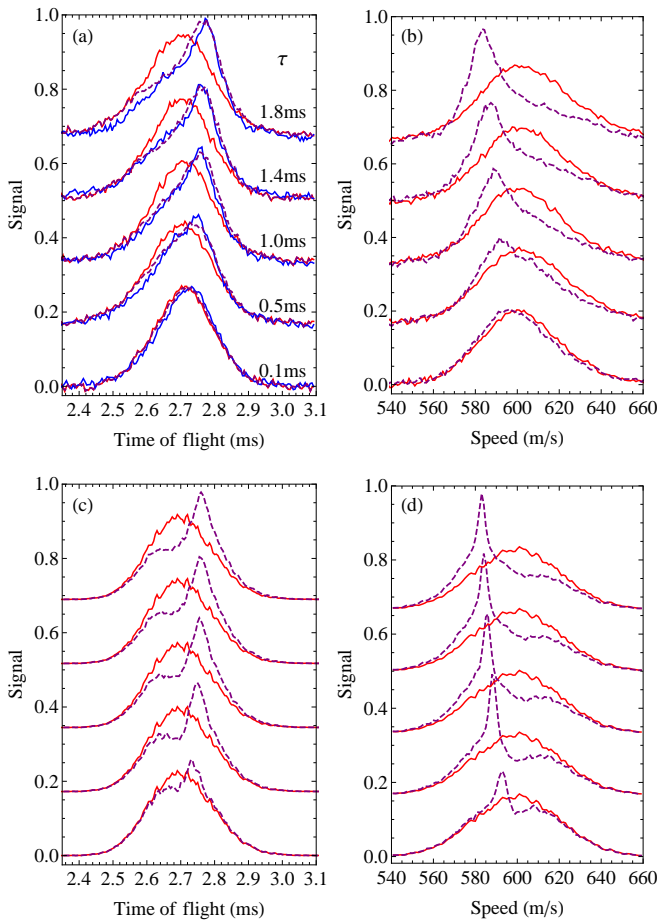}
\caption{(Color online) Effect of laser cooling, without frequency chirp, for various durations, $\tau=0.1,0.5,1.0,1.4$ and 1.8\,ms. (a) Experimental ToF profiles with cooling off (red) and on (blue). Dashed purple lines show the best-fit ToF profiles predicted from the known initial velocity distribution and the simple model for the velocity-dependent force discussed in the text. (b) Velocity profiles inferred from the measured ToF profiles in (a) using this same simple model, with cooling off (solid red) and on (dashed purple). (c) Simulated ToF profiles and (d) velocity profiles, with cooling off (solid red) and on (dashed purple). All profiles are summed over $v''=0$ and $v''=2$ populations.}
\label{VariableCooling}
\end{figure}

Figure \ref{spectrum} shows how the $\text{A} - \text{X} (0-0)$ spectrum changes when laser cooling is applied, with $\tau=1.8$\,ms and $\beta=0$. Some signal is lost because the light optically pumps molecules into $v''=2$, but this reappears in the equivalent $\text{A} - \text{X} (2-2)$ spectrum. In addition, the relative peak heights change because the light re-distributes the population amongst the hyperfine states. To obtain the ToF profile we integrate over this spectrum, applying a different weighting factor to each hyperfine component that accounts for their differing photon yields. These weighting factors are calculated from the spectrum measured without cooling, where we assume that each hyperfine level is populated according to its degeneracy. This procedure ensures that the ToF profiles are not complicated by the effects of velocity-selective hyperfine pumping.

Figure \ref{VariableCooling}(a) shows the ToF profiles when the lasers are tuned to maximize the scattering rate from molecules moving at 600\,m/s.  The symmetrical (red) curves are profiles measured without laser cooling ($\tau=0$), while the other (blue) solid curves are for various nonzero values of $\tau$. In all cases, we sum the populations measured in $v''=0$ and $v''=2$. No population remains in $v''=1$ because the $v_{10}$ light, applied continuously, pumps these molecules to $v''=0$. These profiles show that molecules removed from the original distribution are both slowed and cooled, to produce a peak that is both later and narrower.  The width of the peak together with the relation $v=L/t$, gives an immediate estimate of the velocity distribution and hence of the temperature. This is not accurate because the molecules are decelerated, but it does provide an upper limit on the temperature, which we find to be 430\,mK for the ToF peak at $\tau=1.8\,$ms. The actual temperature is lower, as we now discuss.

Because the source is spatially compact, the ToF profile measured without laser cooling is a direct measure of the velocity distribution produced by the source. Using a simple model for the light force, we can easily convert this initial velocity distribution into the ToF profile expected when the cooling is applied. We approximate the light force as a function of velocity $v$ by $F\exp(-(\tfrac{v-v_0}{\Delta})^{2})$. A single choice of the fitting parameters $F$, $\Delta$ and $v_0$ reproduces all the profiles measured, as shown by the dashed lines in Fig.~\ref{VariableCooling}(a). With $F$, $\Delta$ and $v_0$ now determined, we use the same model to convert the measured ToF profiles into final velocity distributions. These are plotted in Fig.~\ref{VariableCooling}(b). We see that the peaks in the ToF profiles do indeed correspond to a slowing and a narrowing of the velocity distribution. On fitting the $\tau=1.8$\,ms curve in Fig.~\ref{VariableCooling}(b) to the sum of two Gaussians, we find that the final speed of the cooled bunch is $583\pm2\,$m/s and its temperature is $330\pm70\,$mK. The exact form of the light force in our model is not important: any peak-shaped function will do, for example a Lorentzian gives virtually the same final velocity and temperature as the Gaussian. We are therefore confident of our conclusion that these molecules are both decelerated and cooled.

One should consider whether our TOF profiles could be produced by a position-dependent force that longitudinally focusses the molecules onto the detector, rather than a velocity-dependent force that cools the beam. We have ruled this out in two ways. We have modelled the focussing effect of various, suitably-contrived, position-dependent forces, constrained only by the maximum possible scattering rate. Even in the most extreme cases, we find the effect to be far too weak to produce the narrowing we observe. We have also checked experimentally for any position-dependence of the scattering rate by measuring the optical pumping rate into $v''=1$ when the $v_{00}$ light is pulsed on for a short period. These data shows that the scattering rate has exactly the value and velocity-dependence we would expect, and has no significant position dependence. We use a camera to ensure that the $v_{10}$ and $v_{00}$ beams have the same spatial profiles and are well overlapped.

The best fit parameters in our model have the values $v_0=599\pm3$\,m/s, $F=(0.012\pm0.003) h \Gamma/ \lambda$, and $\Delta = 21\,^{+10}_{-4}$\,m/s, all of which are reasonable. The value of $v_0$ is exactly as expected, this being the target value in the experiment. We expect the force to be $P_{\textrm{ex}}  h \Gamma/ \lambda$, where $P_{\textrm{ex}} $ is the probability of being in the excited state. Using a simple formula for excitation in this multi-level system \cite{Tar13a}, taking an average value for the laser intensity, and accounting for the small detunings of some frequencies from their ideal values, we find $P_{\textrm{ex}} \simeq 0.04$. However, a substantial fraction of the decelerated molecules are pumped into the $v''=2$ state, at which point the force turns off. This is responsible for the lower average force that we infer, as discussed further below.  The value for $\Delta$ is somewhat higher than we estimate from the power broadened linewidth of the transition. This can also be caused by optical pumping into $v''=2$ because there is an effective broadening of the transition associated with the saturation of the scattered photon number.

In order to understand the dynamics of the laser cooled molecules in more detail, we have modelled the experiment using a set of 33 coupled rate equations. For each molecule, we solve these equations numerically to follow the populations of the 24 ground-state Zeeman sub-levels and the 4 excited sub-levels, the axial and radial positions and speeds, and the total number of scattered photons. The excitation rate, the damping rate, and the formulae for calculating the branching ratios, are taken from \cite{WallFC}. The excitation rate is summed over the laser frequency components, each with its appropriate intensity and detuning. The laser beam has a Gaussian intensity distribution with parameters as measured in the experiment. The magnetic field, $B$, results in a rotation of the state vector about the field direction at the rate $\omega_{B}=g \mu_{B} B /\hbar$, but this coherent evolution is strongly damped in the experiment because the ground states are coupled to the short-lived excited state. To model this, the rate equations include terms that damp out population difference between states mixed by $B$ at the characteristic rate $\omega_{B}$. These are states of the same $J$ and $F$ that differ in $M_F$ by $\pm 1$. The Franck-Condon factors, $Z_{v'',v'}$ are set to $Z_{0,0}=0.972$ and $Z_{1,0}=1-Z_{0,0}$. The simulation continues until the molecule decays to $v''=2$. This happens after scattering $n$ photons where, for each molecule, $n$ is chosen at random from the probability distribution $Z_{2,0}(1-Z_{2,0})^n$, with $Z_{2,0}=7.84\times 10^{-4}$. We chose $Z_{2,0}$ so that the simulated fraction of molecules pumped into $v''=2$ best matches the experimental observations. The source emits a Gaussian distribution of forward speeds with a mean of 600\,m/s and a temperature of 3.1\,K, while the initial distribution along $z$ is Gaussian with a full width at half maximum (FWHM) of 20\,mm. The initial radial position and speed distributions have widths of 1.2\,mm and 24\,m/s FWHM. The simulation results are insensitive to these values since only molecules with small radial displacements and speeds are detected.

\begin{figure}[tb]
\centering
\includegraphics[width=0.5\textwidth]{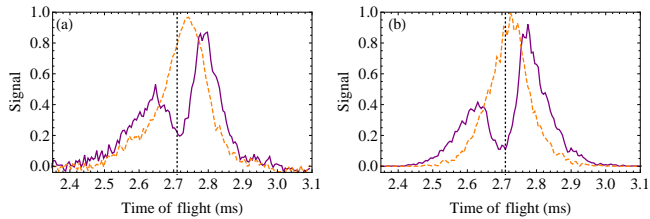}
\caption{(Color online) ToF profiles for molecules in $v''=0$ (solid purple line) and $v''=2$ (dashed orange line) after 1.8\,ms of cooling time without a frequency chirp. Dotted line shows the central arrival time without cooling. (a) Experiment. (b) Simulation.}
\label{PerState}
\end{figure}

Figure \ref{VariableCooling}(c) shows the simulated ToF profiles for our experimental parameters. The arrival time of the cooled peak matches experiment very well in all cases, indicating that the mean force is correctly described by the simulation. The simulated peaks are narrower than in the experiment, and are more prominent than in the experiment for the lower values of $\tau$, suggesting that the cooling is not as strong as the simulation predicts.  The same conclusion follows from a comparison of the simulated velocity distributions, shown in (d), with the distributions inferred from our measurements shown in (b). In the simulation for $\tau=1.8$\,ms, the decelerated peak has a final speed of 583\,m/s, in agreement with experiment, while the final temperature is 85\,mK - somewhat lower than the 330\,mK that we infer from our data.

\begin{figure}[tb]
\centering
\includegraphics[width=0.45\textwidth]{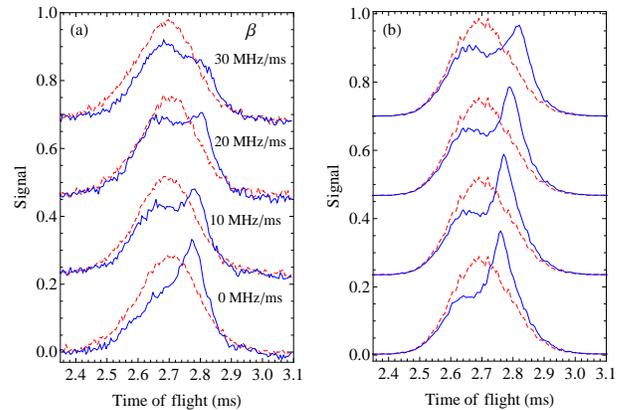}
\caption{(Color online) Effect of laser cooling with various frequency chirps, $\beta=0,10,20$ and 30\,MHz/ms, when $\tau=1.8$\,ms. Dashed red curves: cooling off. Solid blue curves: cooling on. (a) Experimental ToF profiles. (b) Simulated ToF profiles. All profiles are summed over $v''=0,1$ and 2 populations.}
\label{chirping}
\end{figure}

We have investigated the pumping of molecules into $v''=2$ by the cooling lasers. In the experiment, this fraction gradually increases from 6\% of the total when $\tau=0.1$\,ms to 33\% when $\tau=1.8$\,ms. Figure \ref{PerState} shows measured and simulated ToF profiles for molecules in $v''=0$ and $v''=2$ when $\tau=1.8$\,ms. The strong dip in the $v''=0$ profile, close to 2.7\,ms, is due to the interaction with the light. Some of these missing molecules have been decelerated and appear at later arrival times, while the rest have been optically pumped and appear in the $v''=2$ profile. The latter have also been slowed on average, though not by as much as the ones that remain in the cooling cycle for the entire period. All these details are reproduced by the simulation. The dip in the simulated profile is slightly deeper than we measure, and this difference is more pronounced for shorter values of $\tau$ (not shown), however, the arrival times and amplitudes of the peaks in both the $v''=0$ and $v''=2$ profiles match well for all $\tau$.

After $\tau=1.8$\,ms of deceleration, the Doppler shift is so large (28\,MHz or $3.4\Gamma$) that the cooling ceases. This shift can be compensated by chirping the frequencies of the lasers. Figures \ref{chirping}(a) and (b) show the measured and simulated ToF profiles for various chirp rates, $\beta$, with $\tau=1.8$\,ms. (Here, we include the population in $v''=1$ as well as 0 and 2 because the chirp reduces the optical pumping efficiency of the $v_{10}$ laser, particularly at early arrival times). We see that the chirp does indeed slow the molecules further, as intended, and that the delayed arrival times agree well with the simulation results. However, the measured peak height does not agree with  the simulations, being noticeably smaller when the chirp rate is high. We do not yet know what causes this difference. A chirp rate in excess of 30\,MHz/ms produces little further deceleration. This is not surprising because, for $P_{\textrm{ex}} =0.04$, the critical chirp $P_{\textrm{ex}} h \Gamma /(M\lambda^2)$ that exactly matches the changing Doppler shift is 38\,MHz/ms.

In summary, we have shown that the scattering of laser light is able to slow and cool a molecular beam of  CaF molecules. Without chirping, the molecules were slowed by up to 17\,m/s and cooled to 330\,mK. With chirping the deceleration was roughly doubled. Comparison with a detailed numerical model shows that the deceleration is understood, as is the optical pumping into the $v''=2$ state, while the cooling is not quite as strong as predicted.  An additional laser could be used to close the leak to $v''=2$, and then we expect to cool the molecules to the Doppler temperature or below \cite{Shuman(1)09, Tar13a}. Recent advances in doubled fiber laser and diode laser technology \cite{Bal13a} will make this easier to achieve. The cooling presented here would increase the phase-space density of CaF molecules slowed in a Stark decelerator \cite{Wall(1)11}, or a travelling-wave decelerator \cite{Bulleid(1)12} by a factor of 1000. The recent development of cryogenic sources can provide molecular beams with mean speeds of about 50\,m/s \cite{Lu11a}, which could then be laser-slowed to rest and captured in an optical molasses or magneto-optical trap \cite{Stuhl(1)08}.   With its favourable Franck-Condon factors and large electric and magnetic dipole moments, CaF is ideal for exploring the physics of strongly-interacting many-body quantum systems.

This work was supported by the EPSRC and the Royal Society.

\end{document}